\documentstyle[epsf,12pt, bbm]{article}
\newcommand{\be}{\begin{equation}}
\newcommand{\ee}{\end{equation}}
\newcommand{\balpha}{\mbox{\boldmath$\alpha$}}
\newcommand{\ba}{\mbox{\boldmath$a$}}
\newcommand{\bb}{\mbox{\boldmath$b$}}
\expandafter\ifx\csname mathbbm\endcsname\relax
\newcommand{\bzeta}{\mbox{Z\hspace{-0.8em}Z}}
\newcommand{\bone}{\mbox{1\hspace{-0.65em}1}}
\else
\newcommand{\bzeta}{{\mathbbm{Z}}}
\newcommand{\bone}{{\mathbbm{1}}}
\fi
\begin{document}
\begin{titlepage}
\begin{flushleft}
       \hfill                      USITP-95-06, UUITP-4/95\\
       \hfill                       April 1995\\
\end{flushleft}
\vspace*{3mm}
\begin{center}
{\LARGE The Moduli Space and Monodromies of \\
        N=2 Supersymmetric \(SO(2r+1) \) Yang-Mills Theory\\}
\vspace*{12mm}
{\large Ulf H. Danielsson\footnote{E-mail: ulf@rhea.teorfys.uu.se} \\
{\em Institutionen f\"{o}r teoretisk fysik \\

Box 803\\
S-751 08  Uppsala \\
Sweden \/}\\
\vspace*{5mm}
Bo Sundborg\footnote{E-mail: bo@vana.physto.se} \\
        {\em Institute of Theoretical Physics \\
         Fysikum \\
        Box 6730\\
        S-113 85 Stockholm\\
        Sweden\/}\\}
\vspace*{15mm}
\end{center}

\begin{abstract}
We write down the weak-coupling limit of N=2 supersymmetric Yang-Mills
theory with arbitrary gauge group \( G \). We find the weak-coupling
monodromies represented in terms of \( Sp(2r,\bzeta ) \) matrices
depending on
paths closed up to Weyl transformations in the Cartan space of complex
dimension r, the rank of the group. There is a one to one relation
between Weyl orbits of these paths and elements of a generalized braid
group defined from \( G \). We check that these weak-coupling monodromies
behave correctly in limits of the moduli space corresponding to
restrictions to subgroups.
In the case of $SO(2r+1)$ we write down the complex curve representing
the solution of the theory. We show that the curve has the correct
monodromies.

\end{abstract}

\end{titlepage}

\section{Introduction}

After the initial solution of $N=2$ SUSY Yang-Mills theory for $SU(2)$,
\cite{SW},
the extension to $SU(N_{c})$ has been studied in several
papers, [2-4].
In this paper we will work out the details of yet another example:
$SO(2r+1)$.

We should point out that the case of $SO(N_{c})$ has recently been studied
in \cite{SI} for $N=1$ with
matter in the vector representation. With one such matter field, the
number of complex dimensions of the moduli space will always be one. In
$N=1$ language, our case
corresponds to matter in the adjoint representation. In that case, the
number of complex dimensions will be $r$, the rank of the group.

The main result of the paper will be a proposal for a complex curve that
will
describe the moduli-space of $SO(2r+1)$. We will perform several tests to
show that the curve has the correct properties. Among other things, we
will recover the known vacuum structure for a $N=1$ theory. In addition,
we give a uniform description of the weak coupling monodromies for all
simple groups.

In $N=1$ super-space the Lagrangian is
given by
\be
{ 1 \over 4 \pi} Im \left( \int d^{4} \theta { \partial {\cal F} (\Phi )
\over \partial \Phi } \bar{\Phi} + \int d^{2} \theta {1 \over 2}
{ \partial ^{2} {\cal F} ( \Phi ) \over \partial \Phi ^{2}} W^{2}
\right)   ,   \label{verkan}
\ee
where $W = (A,\lambda )$ is the a gauge field multiplet and $\Phi =(\phi , \psi
)$ is a chiral
multiplet, both taking values in the adjoint representation.
The field $\phi$ is given a vacuum expectation-value according to
\be
\phi = \sum _{i=1}^{r} b_{i} H_{i}   ,\label{Cartan}
\ee
where $r$ is the rank of the group. $H_{i}$ are elements of the
Cartan sub-algebra, and we have normalized so that $Tr(H_{i}H_{j})=
\delta _{ij}$.

At a generic $\bb$ the gauge-group is broken down to
$U(1)^{r}$ and each $W$-boson, one for each root $\balpha$, acquires a
$(mass)^2$
proportional to $(\bb \cdot \balpha )^{2}$.
Restoration of symmetry
(classically) is obtained when $\bb$ is orthogonal to a root.
At such a point the $W$-boson corresponding to that root becomes
massless.

The matrix of effective couplings, $\tau ^{ij}$, is given by a one-loop
expression
\be
\tau ^{ij} \sim {\partial ^2 \over \partial b_{i} \partial b_{j}} {\cal F} (\bb
) \sim
{i \over 2\pi} \sum _{\balpha} \alpha ^{i} \alpha ^{j}
\log{ (\bb \cdot \balpha )^2  \over \Lambda ^2} .
\ee
The sum is over all roots $\balpha $ of the algebra. Under a $U(1)_{R}$
transformation $\theta \rightarrow e^{i\omega} \theta$
the field $\phi$ has charge two and hence
\be
\tau ^{ij} \rightarrow \tau ^{ij} - {2\omega \over \pi} \sum _{\balpha} \alpha
^{i} \alpha ^{j}
= \tau ^{ij} - {2\omega C_{2} \over \pi} \delta ^{ij}  ,    \label{coupling}
\ee
where $C_2$ is the eigenvalue of the quadratic Casimir in the adjoint
representation \cite{AKMRV}. From this it follows that the second term in
(\ref{verkan}), i.e.
\be
I = {1 \over 8\pi} Im \left( \int \tau ^{ij} \left( F_{i} F_{j} + i F_{i}^{*}
F_{j} \right) \right) ,
\ee
transforms as
\be
I \rightarrow I - {\omega C_{2} \over 4 \pi ^2} \int  F^{*}F =
I - 4\omega C_{2}n
\ee
where $n$ is the instanton number.
This is consistent with the presence of the ABJ anomaly in the $U(1)_{R}$
current. The $U(1)_{R}$ symmetry is hence
broken down to $\bzeta  _{4C_{2}}$ of which a $\bzeta
_{2C_{2}}$ acts on $\phi$
or $\bb$.

We also need the pre-potential ${\cal F}$ in the semi-classical limit. For a
general group it is obtained
by integrating the effective coupling (\ref{coupling}) twice.
\be
{\cal F} \sim {i \over 4 \pi}
\sum _{\balpha} (\Psi \cdot \balpha )^2 \log{ (\Psi \cdot \balpha )^2  \over
\Lambda ^2}    ,
\label{F}
\ee
where
$\Psi$ is the $N=2$ superfield containing $W$ and  $\Phi$.

\section{Semi-classical monodromies}

When considering the action of \( Sp(2r,\bzeta ) \) monodromy
transformations on the scalar
vevs we use the variables \(a_i\) (defined later in eq.  (\ref{integrals}))
rather than the \(b_i\). In the semi-classical limit they agree and can
be used interchangeably, but in
general they should be distinguished.

The prepotential \( {\cal F} \) is invariant under the discrete Weyl
subgroup of gauge transformations, simply because the sum in eq.
(\ref{F}) is over the set of all roots, which is itself invariant.
However, the logarithms imply that paths starting at a point \( \ba \) and
ending at one
of its Weyl images \( w \ba \) give rise to shifts of \( {\cal F} \) when
the path encircles some singular hyperplanes \( \ba \cdot \balpha = 0 \).
These planes are precisely the walls of the Weyl chambers in the
complexified Cartan
subalgebra. We fix the branch cuts of the logarithms in eq.  (\ref{F}) to
lie along the negative real axis. Monodromy shifts
can then be calculated
by keeping track of how the path encircles the reflection hyperplanes in
passing from Weyl chamber \(C\) to Weyl chamber \(w C\). Each time a
branch cut is passed a shift of \( {\cal F} \) is given by the change in the
imaginary part of the logarithm, compared to the original
expression.

For \( SU(r+1) \) the Weyl group is the permutation group on \(r+1\)
elements and it is generated by the \( r\) reflections \( s_k \) in the
walls of the fundamental domain \(D\) given by
\begin{equation}
	Re( \ba \cdot \balpha _k ) > 0.
	\label{D}
\end{equation}
Here \( (\balpha_k)_{k=1}^r  \) are simple roots of \(SU(r+1)\). Keeping
track of paths means to keep track of how many units of \( \pi \) the
phase of
\( \ba \cdot \balpha \) changes with as its zero is encircled. In terms of
permutations one may think of permuting complex numbers and keeping track
of how they move around each other in the complex plane. Composition of
paths by joining an endpoint to an initial point then gives rise to a
larger group than the Weyl group, the braid group on \( r+1 \) elements.

For other simple groups there is an appropriate generalization of the
ordinary braid group called the Brieskorn braid group, defined solely
in terms of the Weyl group \cite{A}  (or
equivalently the Dynkin diagram of the group). To each vertex of a Dynkin
diagram corresponds a simple root \( \balpha_k \), a Weyl reflection \( s_k
\) and a simple braid \( t_k \). The braid and Weyl groups are generated
by these sets of \(r\) elements and the defining relations
\begin{eqnarray}
	\underbrace{t_m t_n t_m \ldots }_{l_{mn}}
	\underbrace{\ldots  t_n^{-1} t_m^{-1} t_n^{-1} }_{l_{mn}} & = & 1
	\label{braid} \\
	\underbrace{s_m s_n s_m \ldots }_{l_{mn}}
	\underbrace{\ldots  s_n^{-1} s_m^{-1} s_n^{-1} }_{l_{mn}} & = & 1
	\label{Weyl} \\
	s_m^2 & = & 1
	\label{involution}
\end{eqnarray}
where \(l_{mn} \) is $3,4$ or $6$ for $1,2$ or $3$ links, respectively,
joining vertex
\(m\) and \(n\)  of the Dynkin diagram. The braids of the braid group are
in a one to one correspondence with the equivalence classes of paths that
each pick up different logarithmic contributions in moving from a base
point
to one of its Weyl images.

So far we have only discussed how to multiply braids, but we are also
interested in how they act on the physical fields, \( a_i \) and their
duals
\begin{equation}
	 a_D^i \equiv { \partial {\cal F}  \over \partial a_{i}} \sim {i \over 2 \pi}
	\sum _{\balpha} (\ba \cdot \balpha ) \alpha^i
	( 1+ \log{ (\ba \cdot \balpha )^2  \over \Lambda ^2}).
	\label{aD}
\end{equation}
The non-logarithmic term is a Weyl invariant matrix multiplying \( a_i
\), so it has to be proportional to the Cartan metric. We choose an
orthonormal basis of Lie algebra to simplify the notation. (Note that the
bases of the \( SU(r+1) \) Cartan algebras in \cite{AF} and
\cite{KLYT1}
are non-orthonormal). Then the semi-classical monodromies from (\ref{aD})
take the general form
\begin{equation}
	\pmatrix{
\ba_D \cr
\ba\cr
}
\to
		M
	\pmatrix{
	 \ba_D \cr
	 \ba\cr
}
\equiv
	\pmatrix{
w & m \cr
0 & w \cr
}
	\pmatrix{
\ba_D \cr
\ba\cr
}
	\label{class mon}
\end{equation}
where \( m \)
contains the non-trivial
contribution to the monodromy from the logarithms. The discussion of the
Brieskorn braid group applies to the evaluation of \( m \). In
particular, it is enough to calculate the monodromy matrices for
\( r \) generating braids, since the rest of the monodromies should
represent
the braid group. However, some care is needed in order to multiply the
monodromies,
since the minimum number of singularities which a point has to encircle in
going from \(\ba\) to its Weyl image \(w \ba\) depends on which Weyl chamber
\(\ba\) belongs to.

Let us try to generate all semi-classical monodromies from \(r\) simple
monodromies which only
receive contributions from a single
logarithm of (\ref{aD}). This can be achieved by identifying the simple
monodromies with the \(r\) transformations resulting from paths winding
(in the  positive sense) around the \(r\) walls of the Weyl chamber
containing  \(\ba\). For example, \(\ba\) can belong to the fundamental
domain given by eq.  (\ref{D}). Then there is a Weyl reflection \(
s_k \) about each wall \( \ba \cdot \balpha_k = 0 \), and in the
corresponding
monodromy path only the logarithm with argument \((\ba \cdot \balpha_k)^2\)
encounters a branch cut. At the branch cut an
additional phase of \(2 \pi\) is generated, and a non-trivial
monodromy matrix $M_k$ can be read off:
\begin{equation}
	M_k \equiv
	 \pmatrix{
		\bone-2 {\balpha_k \otimes
 \balpha_k \over \balpha_k^2} &   -\balpha_k \otimes \balpha_k \cr
					0 & \bone-2 {\balpha_k \otimes \balpha_k \over \balpha_k^2} \cr
			}   .
	\label{simple mon}
\end{equation}
Even though the form of these simple monodromies is determined by their
action
on \(\ba \) in the fundamental domain \(D\), their domain of definition can
be extended by linearity to the whole set of regular Weyl orbits (the
Cartan subalgebra minus the Weyl chamber walls). Then it is possible to
define the products of simple monodromies simply as matrix
multiplication. However, in Weyl chambers \(C\) other than \(D\) the
\(M_k \) represent paths
through the images of
the walls of \(D\) under the (unique) Weyl
transformation \(w\) mapping \(D\) to $C = w D$, rather than
paths through the walls of the fundamental domain itself.

Then the monodromies \(M_k\) should generate a representation of the
Brieskorn braid group. Indeed, we have checked, for all simple groups,
that they satisfy
\begin{equation}
	\underbrace{M_m M_n M_m \ldots }_{l_{mn}}
	\underbrace{\ldots  M_n^{-1} M_m^{-1} M_n^{-1} }_{l_{mn}}  =  1 ,
	\label{braid rep}
\end{equation}
which are the images of the defining relations  (\ref{braid}) of the braid
group.

\section{The complex curve for $SO(2r+1)$}

\subsection{Constructing the curve}

We will now construct a complex curve appropriate for the $SO(2r+1)$ case. It
must satisfy two requirements.
First it must be symmetric under the Weyl group. Second, it should be
symmetric under the $U(1)_R$ acting on $\phi$ in the (unphysical) limit
of vanishing $\Lambda$, but due to instantons this symmetry should be
broken to $\bzeta _{2C_{2}}$ for general $\Lambda$. For $SO(2r+1)$
we have that
$C_{2} = 2r-1$\footnote{We do not consider $SO(3)$, which is an exception
to the present discussion.}. The Weyl
group of $SO(2r+1)$
acts as the group of all permutations and sign changes of the $r$ numbers
$b_i$ of eq. (\ref{Cartan}). The $b_i$ can be used to parametrize the
curve and have
$U(1)_R$ charge two. With this information it is
natural to try the curve
\be
y^2 = ((x^2 - b_{1}^{2})...(x^2 - b_{r}^{2}))^{2} - \Lambda ^{4r-2} x^2 .
\label{SOcurve}
\ee
If we assign $U(1)_R$ charges $2$ to $x$ and $4r$ to $y$, we see that the
$U(1)_R$ symmetry of the curve is broken to $\bzeta _{4r-2}$ by the
term depending on $\Lambda$. In fact, since the inversion $\bzeta
_{2}$ of the Weyl-group
(as in the case of $SU(2)$) coincides with the $\bzeta  _{2}$ inside
$\bzeta  _{4r-2}$,
there will only be a $\bzeta   _{2r-1}$ acting on the moduli-space.
Furthermore, the curve will depend on $\phi$ only through Weyl invariant
polynomials, for which the Casimirs
$Tr(\phi ^{2k})$ provide a basis.

To verify that this indeed is the correct curve, it
is necessary to check the semi-classical
monodromies of the curve.

\subsection{Partial symmetry breaking and factorization of the curve}

The monodromies can be obtained from the curve by introducing \(r\)
pairs of homology cycles \( (\gamma_D^i,\gamma_i) \) on
the curve, one for each \(b_i\), and then following how they transform as
the
parameters of the curve are varied to a Weyl equivalent point.
The semi-classical limits consist in taking some
\begin{equation}
	|b_i| >> \Lambda .
	\label{limit}
\end{equation}
The fields \( a_i \) will be defined in
terms of the curve so as to ensure that \(b_i \to a_i\) smoothly in
these  limits. Then a transformation of the homology cycles \(
(\gamma_D^i, \gamma_i) \) can be interpreted as a monodromy transformation
on the \( (a_D^i, a_i)^t \) vectors. By taking only a subset of the \(
b_i \) to be large, the symmetry is only partially broken. Hence, it
should be possible to read off the curves corresponding
to subgroups of the full unbroken gauge symmetry by taking appropriate
limits of the \(b_i\). For the \(SO(2r+1)\) curves (\ref{SOcurve}) it
works as
follows.

By deleting a vertex from the Dynkin diagram of a group, one gets
subgroups of rank \(r-1\). These subgroups are seen from the curve by
taking the limit
\( \bb \cdot \balpha_k \to \infty \) while keeping the scalar products with
the other simple roots fixed, i.e. by moving \( a \) far
away from the reflection hyperplane of the root \( \balpha_k
\) corresponding to the  deleted vertex. For \(SO(2r+1)\) a
canonical choice of simple roots is
\begin{eqnarray}
	\balpha_k & = & \hat{e}_k-\hat{e}_{k+1}, k=1, \ldots , r-1
	\label{suroots} \\
	\balpha_r & = & \hat{e}_r
	\label{soroot}
\end{eqnarray}
in terms of an ON basis. Deleting \(\balpha_1\) produces simple roots of
\(SO(2r-1)\), deleting \(\balpha_r\) gives simple roots of
\(SU(r)\) and  removing one of the other roots \(\balpha_k\)
yields \(SU(k)\times  SO(2r-2k+1)\). The prepotential \( {\cal F}\)
respects this sub-group  structure and so should the curve
(\ref{SOcurve}). Let us check the case when a vertex inside the Dynkin
diagram is deleted.

Taking the limit \( \bb \cdot \balpha_k \to \infty \) and defining new
parameters
\begin{eqnarray}
	k B & = & b_1 + \ldots + b_k
	\label{B} \\
	\hat{b}_i & = & b_i - B, i=1, \ldots, k
	\label{b'}
\end{eqnarray}
the curve takes the form
\begin{eqnarray}
	y^2 & = & (x - \hat{b}_{1}-B)^2 \ldots (x - \hat{b}_{k}-B)^2
	(x +\hat{b}_{1}+B)^2 \ldots (x + \hat{b}_{k}+B)^2  \nonumber  \\
	& & \times (x^2 - b_{k+1}^{2})^2 \ldots (x^2 - b_{r}^{2})^2 - \Lambda ^{4r-2}
x^2 .
	\label{factorcurve}
\end{eqnarray}
We see that the product on the right hand side of the equation factorizes
into three groups of factors. In the limit \(B\to\infty\), which is
precisely the limit we are considering, the zeroes inside each group are
at finite distances from each other, while distances between zeroes in
different groups diverge. For fixed \(x\) the curve then approaches the form
\begin{equation}
	y^2 = (x^2 - b_{k+1}^{2})^2 \ldots (x^2 - b_{r}^{2})^2
	- \Lambda ^{4r-4k-2} x^2
	\label{solimit}
\end{equation}
after a rescaling of \(y\) and renormalization group
matching\footnote{Again $SO(3)$ for $k=r-1$ is
an exception to the discussion.} of
\(\Lambda_{SO(2r-2k+1)}\) to \(\Lambda_{SO(2r+1)}\). We
recognize the expression for an \(SO(2r-2k+1)\) curve!

If instead the regions around \(x= \pm B\) are studied, one
obtains in a similar way
\begin{equation}
	y^2 = ((x \mp \hat{b}_{1}) \ldots (x \mp \hat{b}_{k})) ^2
	- \Lambda^{2k}
	\label{sulimit}
\end{equation}
after also shifting \(x\). These curves are precisely the \(SU(k)\) curves
of
Argyres and Faraggi \cite{AF} and of Klemm, Lerche, Yankielowicz and
Theisen \cite{KLYT1}. Even though two \(SU(k)\) curves appear in
this factorization they only represent one group factor, since they are
exact mirror images of each other. For removal of vertices at the ends of
the Dynkin diagram, \(k=1\) or \(k=r\), almost identical arguments give
the expected symmetry breaking.

\subsection{Checking the monodromies for $SO(5)$}

Let us consider the $SO(5)$ case in some detail!
The semi-classical monodromies can be checked by acting with the braid group
on the $b$'s and keeping track of the cycles. We choose a basis
of cycles as in figure 1.

\begin{figure}

\hbox to\hsize{\hss
\epsfysize=8cm
\epsffile{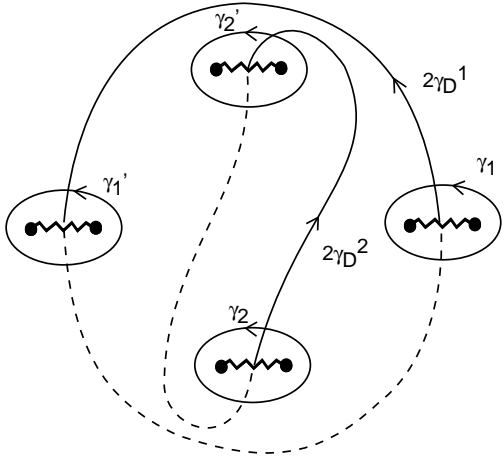}\hss}
\caption{Cycles for the SO(5) curve.}
\end{figure}

The fields transforming simply under \( Sp(2r,\bzeta ) \) duality
are $a_{i}$ and $a_{D}^{i}$, and they are given by the
following integrals over cycles
\be
a_{i} = \oint _{\gamma _{i}} \lambda \;\;\;\; \mbox{and} \;\;\;\;
a_{D}^{i} = \oint _{\gamma _{D} ^{i}} \lambda  . \label{integrals}
\ee
Let us derive the expression for
the one-form $\lambda$ in the case of a curve of
the more general form
\be
y^2 = p^2 (x) -x^k \Lambda ^m
\ee
where $k<2n$ and $p(x)$ is a polynomial of order $n$ of the form
$ p(x) = x^n +\sum _{i=1}^{n-2} u_{i} x^i$.
The genus of
such a curve is $n-1$. There are $n-1$ holomorphic one-forms
${dx \over y},..., { x^{n-2} dx \over y}$ and
$n-1$ meromorphic one-forms ${ x^n dx \over y} ,..., { x^{2n-2} dx \over
y}$,
with vanishing residue. $\lambda$ must be a combination of these one-forms
up to exact pieces.
Furthermore, all derivatives ${\partial \lambda \over
\partial u_{i}}$ must consist of holomorphic one-forms only.
These requirements imply that
\be
\lambda = ({k \over 2 } p - x p') {dx \over y}  .
\ee
For $k=0$ we recover the expression for $SU(n)$, while for $k=2$ we
find
\be
\lambda = (p- xp') { dx \over y}
\ee
with $p(x) = (x^2-b_{1}^2)...(x^2-b_{r}^{2})$ appropriate for $SO(2r+1)$.

Due to the reflection symmetry of the curve, $x \rightarrow -x$, which
$\lambda$ respects,
we have
\be
\oint _{\gamma _{i}} \lambda = - \oint _{\gamma '_{i}} \lambda .
\ee
The presence of this symmetry
is a complication not present in the case of $SU(n)$. In fact, the genus
of the curve for $SO(2r+1)$ that we propose is $g=2r-1$ while the rank of
the group is just $r$.
However, due to the symmetry the curve is described by just $r$ parameters.

The effect of the Weyl transformations on the branch-points are shown in
figure 2.
\begin{figure}
\hbox to\hsize{\hss
\epsfysize=8cm
\epsffile{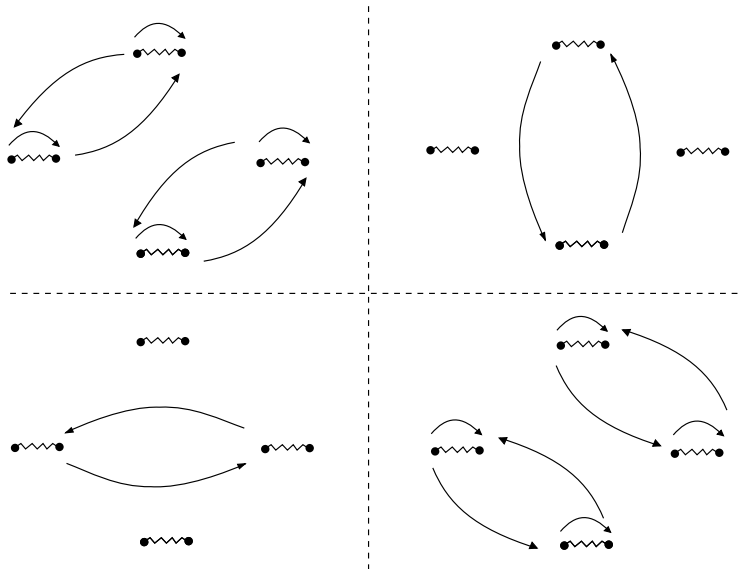}\hss}
\caption{The action of  the braidings on the branch-points
for SO(5).}
\end{figure}
Represented on $(a_{D}^{i} , a_{i})$, using the reflection
symmetry, the monodromy matrices corresponding to the braidings in the figure
can be found to be
\begin{eqnarray}
M_{1} = \pmatrix{
0 & 1 & -1 & 1 \cr
1 & 0 & 1 & -1 \cr
0 & 0 & 0 & 1 \cr
0 & 0 & 1 & 0 \cr
}
 , &
M_{2} =\pmatrix{
1 & 0 & 0 & 0 \cr
0 & -1 & 0 & -1 \cr
0 & 0 & 1 & 0 \cr
0 & 0 & 0 & -1 \cr
} , \nonumber \\
M_{1} M_{2} M_{1}^{-1} = \pmatrix{
-1 & 0 & -1 & -2 \cr
0 & 1 & 2 & 0 \cr
0 & 0 & -1 & 0 \cr
0 & 0 & 0 & 1 \cr
}
 , &
M_{2}^{-1} M_{1} M_{2} = \pmatrix{
0 & -1 & -1 & -2 \cr
-1 & 0 & 0 & -1 \cr
0 & 0 & 0 & -1 \cr
0 & 0 & -1 & 0 \cr
} ,
\label{klassmon}
\end{eqnarray}
respectively.
We must now now compare with the result obtained from working with the
semi-classical action.
It is easy to see that the results agree. $M_{1}$ and $M_{2}$ are simple
monodromies of the form (\ref{simple mon}).

\subsection{The special Coxeter monodromies}

Argyres and Faraggi \cite{AF} used special monodromies to check the
semi-classical monodromies of \(SU(n)\) curves. To use their arguments in
a
more  general context we rephrase them as follows.

We have noted that all
semi-classical monodromies can be generated by
composition of \(r\) simple monodromies $M_k$.
Therefore, when all monodromies of the rank \(r-1\)
sub-groups obtained by removing a vertex from a rank \(r\) Dynkin
diagram are known, it is enough to check a single new
semi-classical monodromy for the rank \(r\) group. The simplest
monodromies which do not belong to
any such sub-groups correspond to braids
\begin{equation}
	t_C = t_{1} t_{2} \ldots t_{r}.
	\label{Coxeterbraid}
\end{equation}
One can prove that the effect
of a permutation of the factors of the product is only to give a braid
group element which is conjugate to \( t_C \). The argument is a copy of
a standard argument \cite{H} for the corresponding element in the Weyl
group, the Coxeter element. We therefore call \( t_C \) a Coxeter braid.

We have already checked that the curve (\ref{SOcurve}) gives the
correct hierarchy of subgroups, and that it works for low rank groups. We
can then prove by induction that composition of semi-classical
monodromies works properly,
if only the special rank \(r\) monodromy corresponding to
the Coxeter braid is given correctly by the curve.

A great advantage of using Coxeter braids is that their monodromies are
easy to calculate from the curve by perturbation theory in small
\(\Lambda\). The \(SO(2r+1)\) curve (\ref{SOcurve}) has branch points at
the roots of the right-hand side polynomial in \(x\). For \(\Lambda = 0\)
there are \(2r\) double roots at \(x = \pm b_k\). Let \(b_{1}\) be a
complex number such that \(0<{\rm Arg}b_{1}<\pi / 2r\), and set
\begin{equation}
	b_{k+1} = b_{k} e^{i k (-1)^k \pi / r}, k = 1, \ldots , r - 1.
	\label{b}
\end{equation}
Then the vector \(\bb\) lies in the fundamental domain, and our previous
results can be applied to monodromies with \(\bb\) as a base point. For
small non-zero \(\Lambda\) the double roots on a circle split into \(2r\)
pairs of branch points at
\begin{equation}
	b^{\pm}_{k} \approx b_{k} (1 \pm c ({\Lambda \over b_{k}})^{2r-1}),
	\label{splitb}
\end{equation}
where \(c\) is a real-valued constant. It is convenient to choose a basis
of cycles analogous to the $SO(5)$ cycles of figure 1, with cuts between
the branch points of a pair, $\gamma$ and $\gamma'$ cycles around the
branch cuts and $\gamma_D$ cycles between opposite branch cuts.

One finds that a Coxeter braid $t_{r}t_{r-2} \ldots t_{r-3}t_{r-1}$
rotates the positions of pairs $b_k$ an angle $\pi/r$ while the cuts
are rotated an angle $2\pi - 2\pi/r$ in the opposite direction. From this
we have checked directly the special monodromies of $SO(5)$ and $SO(7)$,
but it is much easier to check their $2r$'th power.
Then all $b$'s are
rotated a full turn and their cuts rotate $2r-2$ times in the opposite
direction. This will cause a $2\gamma _{D}^i$ cycle to wind $2(2r-2+1)$
times around $\gamma _{i}$ and the same number of times around $\gamma
_{i}'$. Each sheet is contributing, hence the factor of $2$. From this it
follows that this power of the special monodromy is given by
\be
\pmatrix{
\bone & -2C_{2}\bone  \cr
0 & \bone   \cr }
\ee
for any $SO(2r+1)$ group, as is to be expected from (\ref{aD}).

\section{Strong coupling monodromies}

Let us now consider the singular sub-manifolds of the moduli space
where the discriminant vanishes. Parametrized by  the symmetric polynomials $u$
and $v$, the $SO(5)$ curve is given by
\be
y^2 = (x^4 - ux^2 +v)^2 - \Lambda ^{6} x^2    .
\ee
The discriminant vanishes when $v=0$ or for $u$ and $v$ such that
\begin{eqnarray}
x^4 - u x^2 +v & = & \pm \Lambda ^{3} x \nonumber \\
x ( 4 x^2 - 2u) & = & \pm \Lambda ^{3}
\end{eqnarray}
has a solution. One may study, for instance, the three dimensional submanifold
$Im(v)=0$.
All $Re(v) = const.$ planes (if $Re(v) \neq 0$) can be shown to cut through
four
singular submanifolds. As $Re(v)= 0$ is approached, one of these singular
submanifolds will drift of towards $|u| = \infty$, and only three will cut
the $Re(v)=0$ plane. They will do so in the points
\be
u= { 3 \over 4^{1/3} } \Lambda ^{2} e ^{{2 \pi i \over 3 } n}
\ee
where  $n=1,2,3$ respectively. One can check that each intersection-point
corresponds to  a pair of mutually local dyons becoming massless. The three
intersection-points
are related by a $\bzeta _{3}$-symmetry and
correspond to the three vacua of a  $N=1$ $SO(5)$ theory where we have
confinement or oblique confinement.  That these vacua are represented in the
$N=2$ theory, is an important check of our
curve. We might add that there is another triplet of intersections that
are not candidates for $N=1$ vacua. These are at
\be
u = {3\Lambda ^2 \over 2} e^{{2\pi i \over 3 } n},\;\;\;
v = -{3 \Lambda ^4 \over 16} e^{{4\pi i \over 3 } n}
\ee
for $n=1,2,3$.

A strong coupling monodromy is
obtained by encircling some singular sub-manifold.
The semi-classical monodromies can often be obtained by taking
pairs of such strong coupling monodromies. A strong coupling monodromy
typically splits some pair of branch points. The semi-classical
monodromies, as
we have seen, leave the pairs together. Let us consider an example!

To find the strong coupling monodromies one can use the Picard-
Lefshetz theorem. This is based on the vanishing cycles and states
that
${\cal M} _{\nu} \gamma = \gamma + \langle \gamma , \nu \rangle \nu  $
where $\gamma$ is an arbitrary cycle on which the monodromy ${\cal M}
_{\nu}$ is acting
and $\nu$ the vanishing cycle.
A vanishing cycle is a cycle that encircles a pair of branch points
that move together as some singular sub-manifold is approached.
A pair of such vanishing cycles are shown in figure 3.

\begin{figure}
\hbox to\hsize{\hss
\epsfysize=7cm
\epsffile{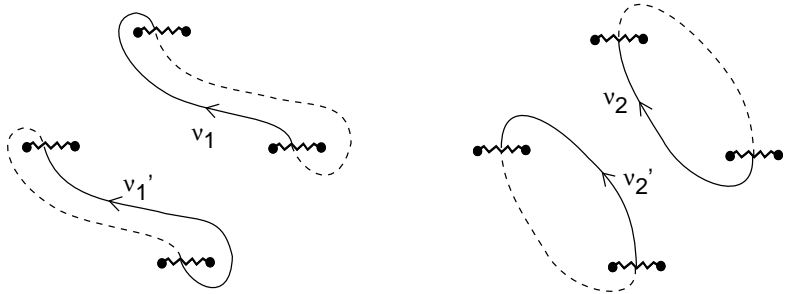}\hss}
\caption{Vanishing cycles for two strong coupling monodromies for
SO(5).}
\end{figure}

Note that
due to the reflection symmetry, each vanishing cycle is doubled.
We hence  have to be careful when we use the Picard-Lefshetz theorem.
In fact, we have
\be
{\cal M} _{\nu } \gamma = \gamma + \langle \gamma , \nu \rangle \nu +
\langle \gamma , \nu ' \rangle \nu '.
\ee
Alternatively one can trace the cycles as the branch points move as
we did for the semi-classical case.
The strong coupling monodromies so obtained are
\be
{\cal M} _{1} =
\pmatrix{
0 & -1 & -1 & -2 \cr
-2 & -1 & -2 & -4 \cr
1 & 1 & 2 & 2 \cr
1 & 1 & 1 & 3 \cr
}
\ee
and
\be
{\cal M} _{2} =
\pmatrix{
1 & 0 & 0 & 0 \cr
-1 & 0 & 0 & -1 \cr
1 & 1 & 1 & 1 \cr
1 & 1 & 0 & 2 \cr
}     .
\ee
Multiplying them together as ${\cal M} _{2} {\cal M} _{1}$ gives the last
of the
monodromies in eq.  (\ref{klassmon}) as it should.
It is easy to check that ${\cal M} _{1}$ is associated with a dyon with
charge vector
$(\bar{g}_{1},\bar{q}_{1}) = (1,-1,1,-2)$
and ${\cal M} _{2}$ with a dyon with charge vector
$(\bar{g}_{2},\bar{q}_{2}) =(1,-1,0,-1)$. This can be read off directly from
the corresponding vanishing cycle or by noting that these charge-vectors
are left eigen-vectors
with eigen-value one for the respective monodromy matrice.

Similarly, one can work out the details for the other singular
sub-manifolds and dyons.

\section{Conclusions}

In this paper we have extended the construction for $SU(N_{c})$ treated in
[1-4] to the case $SO(2r+1)$.  A new and perhaps unexpected feature of the
complex curves involved in our solutions is that the genus is larger than the
rank of the group. This is possible because of additional symmetries of the
curves. Clearly it is important to generalize the
solutions to arbitrary groups. We believe that our general description of the
semi-classical monodromies can be of help in such constructions.

\bigskip

\begin{flushleft}
We wish to thank T. Ekedahl, J. Kalkkinen, U. Lindstr\"om, and H.
Rubinstein for their comments.
\end{flushleft}

\end{document}